\documentclass[twocolumn]{aastex7}

\shorttitle{Phase-Resolved Cross-Correlation Retrievals}
\shortauthors{Savel et al.}
\usepackage{amsmath}
\usepackage{booktabs}
\begin{document}

\title{Precise Determination of the Metallicity and C/O of WASP-39~b From a Single JWST Instrument Mode with Phase-Resolved Cross-Correlation Retrievals}

\correspondingauthor{Arjun B. Savel}
\email{asavel@umd.edu}

\newcommand\umd{\affiliation{Astronomy Department, University of Maryland, College Park, 4296 Stadium Dr., College Park, MD 20742 USA}}

\newcommand\montreal{\affiliation{Institut Trottier de Recherche sur les Exoplanètes, Université de Montréal, Montréal, Québec, H3T 1J4, Canada}}

\newcommand\bern{\affiliation{Weltraumforschung und Planetologie, Physikalisches Institut, University of Bern, Gesellschaftsstrasse 6, 3012 Bern, Switzerland}}

\newcommand\stsci{\affiliation{Space Telescope Science Institute, Baltimore, MD, USA}}

\newcommand\hopkinsapl{\affiliation{Johns Hopkins APL, 11100 Johns Hopkins Road, Laurel, MD 20723, USA}}

\newcommand\hopkins{\affiliation{Department of Earth and Planetary Sciences, Johns Hopkins University, Baltimore, MD 21218, USA}}

\newcommand\uchicago{\affiliation{Department of Astronomy \& Astrophysics, University of Chicago, Chicago, IL 60637, USA}}

\newcommand\asu{\affiliation{School of Earth and Space Exploration, Arizona State University, PO Box 871404, Tempe, AZ 85281, USA}}

\author[0000-0002-2454-768X]{Arjun B. Savel}\email{asavel@umd.edu}
\umd

\author[0000-0002-1337-9051]{Eliza M.-R. Kempton}\email{ekempton@uchicago.edu}
\uchicago
\umd

\author[0000-0002-2739-1465]{Erin M. May}\email{erin.may@jhuapl.edu}
\hopkinsapl

\author[0000-0001-8236-5553]{Matthew C. Nixon}\email{matthewnixon@asu.edu}
\asu
\umd

\author[0000-0003-2775-653X]{Jegug Ih}\email{jih@stsci.edu}
\stsci

\author[0000-0002-9030-0132]{Katherine A. Bennett}\email{kbenne50@jhu.edu}
\hopkins

% Added middle intial and new affiliation
\author[0000-0003-3191-2486]{Joost P. Wardenier}\email{joost.wardenier@unibe.ch}
\bern



\begin{abstract}

Measuring atmospheric metallicities and C/O ratios is a key goal of JWST exoplanet science, given their proposed link to planet formation. Achieving this goal has previously been shown to require broad wavelength coverage ($\sim$1--5 $\mu$m), typically demanding multiple instrument modes to complete the molecular inventory. Here, we show that the multi-instrument requirement can be circumvented using phase-resolved cross-correlation retrievals at native pixel resolution --- an approach more typically applied to ground-based high-resolution spectroscopy. By applying this novel analysis technique to an archival single-mode transit of the hot Jupiter WASP-39~b obtained with NIRSpec/G395H, we detect and obtain bounded abundances for all of its major carbon- and oxygen-bearing molecules: H$_2$O ($\Delta \ln(Z) = 67$), CO ($\Delta \ln(Z) = 25$), CO$_2$ ($\Delta \ln(Z) = 475$), and SO$_2$ ($\Delta \ln(Z) = 10$). Notably, while standard retrieval methods fail to detect CO in these same data ($\Delta \ln(Z) = 0.2$), our approach detects it decisively, confirming it as the dominant carbon carrier in WASP-39 b's atmosphere. From these abundances, we robustly derive a metallicity of $\rm [(C + O) / H] = 1.2 \pm 0.2$ and a C/O ratio of $0.68^{+0.10}_{-0.14} $, generally consistent with previous multi-instrument analyses. In comparison, traditional retrievals performed on the G395H data alone produce biased and inaccurate values of both parameters, driven primarily by the non-detection of CO as well as incomplete water band coverage. Our results establish phase-resolved cross-correlation retrievals as a powerful tool for extracting maximum atmospheric information from existing and future JWST data sets.

\end{abstract}


\section{Introduction} \label{sec:intro}
JWST \citep{gardner2006james} is already delivering on one of its promises: to provide impactful, information-rich measurements of transiting exoplanet atmospheres \citep[e.g.,][]{early2022identification, feinstein2023early,grant2023jwst,bell2024nightside}. In high signal-to-noise JWST transmission spectra, detections of spectral features arising from atmospheric gases are incontrovertible \citep{jwst2023identification, coulombe2023broadband}, revealing exoplanetary processes in unprecedented detail \citep{tsai2023photochemically, challener2024latitudinal}.  Underpinning this success is JWST's capacity to detect a wide variety of molecules, many of which were not accessible with previous facilities, as well as its ability to robustly constrain both atmospheric metallicities ([M/H]) and carbon-to-oxygen ratios (C/O).

Measurements of metallicity and C/O with JWST have been highly anticipated because of their purported role as signposts of planet formation \citep{oberg2011effects,madhusudhan2012c}. Planets' assembly in relation to snowlines in protoplanetary disks should have imprints on their composition, especially via abundance ratios \citep[e.g.,][]{piso2016role,hobbs2022molecular}. Accurately constraining both [M/H] and C/O requires a full spectroscopic accounting of the key carbon and oxygen reservoirs in exoplanet atmospheres (typically H$_2$O, CO$_2$, CO, and CH$_4$ for hot and warm jovians).  Thus, the most informative JWST datasets typically combine spectra from multiple observing modes to compile the full molecular inventory. The primary advantage of the multi-instrument approach is that broader wavelength coverage can include contributions from multiple gaseous absorbers, some with multiple absorption bands observed, which aids in breaking degeneracies between molecular abundances and aerosol properties \citep[e.g.,][]{benneke2013}.  Both simulations \citep{batalha2017information} and observations \citep{heinke2026information} have shown that the combination of the NIRISS/SOSS and NIRSpec/G395H observing modes, spanning \mbox{$\sim$0.5--5 $\mu$m}, provides the best constraints on [M/H] and C/O. Single-mode observations, lacking such spectral range, can yield constraints on atmospheric chemistry that are otherwise surprising and difficult to reconcile with expectations from planet formation theory \citep{bean2023high, meech2025bowie, claringbold2026bowie}.

In this Letter, we challenge the idea that multiple observing modes are required to produce robust measurements of [M/H] and C/O by leveraging novel analysis techniques.
At the same time, we are able to address another outstanding concern with JWST data analysis related to the detectability of the CO molecule in exoplanet atmospheres, which we illustrate through the example of WASP-39~b \citep{faedi2011wasp}. 
CO was detected at high significance (7$\sigma$) in WASP-39~b's  NIRSpec/PRISM transmission spectrum \citep{rustamkulov2022analysis}, in line with predictions from equilibrium chemistry \citep[e.g.,][]{stock2018fastchem}. Yet it was not found in the higher spectral resolution NIRSpec/G395H data of the same planet \citep{alderson2023early}, presenting a puzzle.  

The solution is found by considering the behavior of the fundamental CO band at 4.6 $\mu$m.  In typical binned low-resolution JWST data, the band appears wide and weak, despite the intrinsic strength of the individual ro-vibration lines that comprise the feature.  
In two follow-up efforts, \cite{grant2023detection} and \cite{esparza2023detection} both showed that CO can be detected in the same G395H data that produced the \citet{alderson2023early} non-detection, but at \textit{native spectral resolution}. These detections were both enabled by specialized analysis techniques tailored to detect the high-resolution peaks in the CO opacity function. When unbinned, the band structure of CO provides additional spectral correlation that can be distinguished from noise. The mystery is therefore resolved --- CO is indeed present in WASP-39~b's atmosphere, but standard analysis techniques making use of \textit{binned} G395H data do not provide enough sensitivity to the molecule. The implicit tradeoff is that while binned G395H data could not reach the S/N of PRISM to constrain the CO pseudocontinuum, the lower-S/N, higher-resolution unbinned G395H data were more sensitive to the strong and regular wavelength dependence of CO. 

The difficulty of detecting CO is not limited to WASP-39~b.  Few papers have claimed detections of CO in hot Jupiters using JWST data \citep[e.g.,][]{fu2024hydrogen,gapp2025wasp,meech2025bowie}. Even fewer have robust, bounded constraints on the abundance of CO. This paucity is despite the fact that CO is expected to be the dominant carbon carrier in hot Jupiter atmospheres \citep{moses2011disequilibrium}. 

We offer a way around this lack of sensitivity, taking particular interest in the \cite{esparza2023detection} approach: cross-correlation. The authors detected CO at 6$\sigma$ with this technique, with the significance increasing to 7.5$\sigma$ when including minor CO isotopologues. Follow-up work demonstrated that the cross-correlation approach can also be successfully applied to other molecules, detecting an inventory of molecules consistent with standard analyses \citep[i.e., H$_2$O and CO$_2$;][]{esparza2025testing}. These results imply that existing JWST datasets genuinely contain additional signal that traditional techniques are missing. 

The \cite{esparza2023detection} method, while innovative, is limited to molecular \textit{detection}; it does not measure the abundances of gases or quantify their uncertainties. Doing so would maximize quantitative constraints on atmospheric chemistry --- and through chemistry, potentially planet formation \citep[e.g.,][]{oberg2011effects, molliere2022interpreting,feinstein2025linking}. Fortunately, the community of ground-based exoplanet atmosphere observers has spent the last decade addressing a similar problem of extracting the signatures of unseen absorbers in high-resolution spectra and quantifying their abundances \citep[e.g.,][]{brogi2019retrieving, gibson2020detection}. By mapping the cross-correlation to a likelihood function, models and data can be compared in a Bayesian framework, connecting cross-correlation techniques to the well-established discipline of atmospheric retrievals \citep{madhusudhan2009temperature, benneke2012atmospheric}. This approach has the added benefit of explicitly revealing degeneracies between molecules and aliases that could otherwise hide in cross-correlation.

Another point of note is that most analyses of JWST transmission spectra to-date, including the cross-correlation studies of WASP-39~b by \citealt{esparza2023detection}, have made use of the \textit{time-integrated} transit.  That is, the analyses are performed by first fitting the full \mbox{U-shaped} transit to extract a single value --- the transit depth --- at each wavelength.  Subsequent interpretation then occurs on this time-integrated spectrum, which is typically referred to as ``the'' transmission spectrum.  Ground-based high-resolution studies of exoplanet atmospheres typically follow a different approach, in which the \textit{phase-resolved} data are maintained throughout the analysis and are fitted directly \citep[e.g.,][]{prinoth2023time,gandhi2023retrieval,boldt2025vlt}. The reason for this is that the cross-correlation procedure detects exoplanets' signals in velocity space.  As hot Jupiters such as WASP-39~b transit, their projected orbital velocities shift by $\sim$10~km\,s$^{-1}$, which is a substantial fraction of the G395H pixel scale along the spectral axis (74~km\,s$^{-1}$ at native pixel resolution).  It is plausible, then, that (effectively) integrating over the time axis may smear out signal that would otherwise be accessible in phase-resolved spectra with cross-correlation-based methods. Recent phase curve analyses \citep{sing2024absolute, ouyang2026cross,2026arXiv260700952D,snellen2026cancri} imply that this notion may hold for JWST data, as well.

This Letter presents the first phase-resolved cross-correlation retrieval on JWST transmission spectra.  We apply a modified version of techniques previously performed only on ground-based exoplanet spectra to analyze the NIRSpec/G395H spectrum of WASP-39~b.  By focusing on the G395H instrument mode, we endeavor to accurately and precisely measure both [M/H] and C/O for this benchmark hot Jupiter.  This aim is possible because all of the major C- and O-bearing molecules provide considerable absorption across the G395 bandpass, and the cross-correlation approach has improved ability to resolve degenerate solutions through its sensitivity to the many individual ro-vibration lines that comprise each molecular band.  This is in comparison to traditional low-resolution techniques, which for example only fit for a \textit{partial} water absorption band in NIRSpec/G395 measurements, leading to the possibility of inaccurate abundance inference due to oversimplified cloud parameterizations or detector edge effects.

The remainder of this Letter is structured as follows. Section~\ref{sec:methods} describes our methodology, with particular focus on the different steps we take with respect to ``traditional'' JWST retrievals. In Section~\ref{sec:results}, we apply our techniques on archival NIRSpec/G395H observations of WASP-39~b \citep{alderson2023early}.
We discuss the robustness of these results and their implications in Section~\ref{sec:discussion}. Finally, we conclude and remark on future directions in Section~\ref{sec:conclusion}.

\section{Methods} \label{sec:methods}
\begin{figure*}
    \centering
\includegraphics[
    width=0.91\linewidth,
    trim=0 1.5cm 0 0,
    clip
]{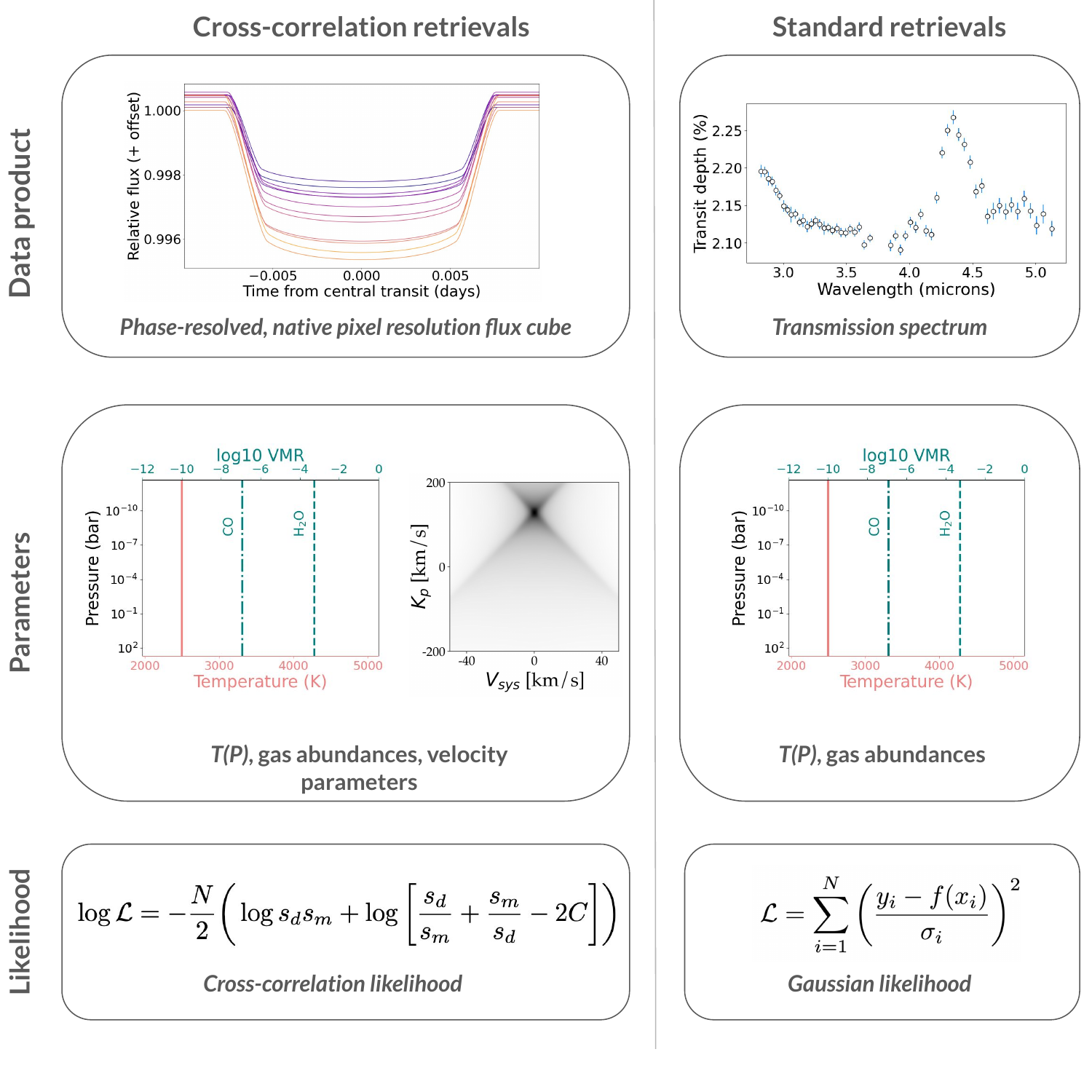}
    \caption{Schematic comparing our retrievals (left) and standard JWST retrieval methodology (right). The retrievals that we present in this work begin with a different data product and fit for additional parameters with a different likelihood function.}
    \label{fig:schematic}
\end{figure*}

Our analysis (depicted in Fig.~\ref{fig:schematic}), as is standard for atmospheric retrievals, has four components: spectral data, a forward atmospheric model, a likelihood function, and a sampling algorithm. For completeness, we review these components as they are implemented in this study, with special attention to our departures from standard analyses of JWST transmission spectra.

\subsection{The data}\label{sec:thedata}
We elect to perform our analysis on the WASP-39~b G395H spectrum (JWST ERS 1366, PI Batalha) previously published in \citet{alderson2023early}. The NIRSpec/G395H observing mode has the highest spectral resolution of the modes commonly used for transiting exoplanet science, and this is the only JWST transmission spectrum with published cross-correlation detections to date \citep{esparza2023detection, esparza2025testing}. We begin by re-reducing the WASP-39~b G395H data, as there have been changes to the JWST data calibration files since the original ERS data reductions. We use the \texttt{Eureka!} pipeline \citep{bell2022eureka}, which serves as a wrapper for the \texttt{jwst} pipeline \citep{bushouse2023} in Stages 1 and 2, with the addition of an extra step to perform group-level background subtraction prior to ramp fitting. This step removes $1/f$ noise due to detector readout and increases the accuracy of ramp fitting.\footnote{\texttt{Eureka!}\ \texttt{.ecf} files used to generate these data will be made available upon publication.}

Corrections to JWST's barycentric motion are applied during the data reduction process when setting the common wavelength solution for our spectroscopic lightcurves in Stage 2. We do not account for systemic velocity or the orbital motion of the planet in the data reduction stage. Therefore, the velocity offset that we expect is simply the system's velocity in the barycentric frame: $-58.4421$~km\,s$^{-1}$ \citep{mancini2018gaps}. We concatenate the data from both detectors (NRS1 and NRS2), resulting in a single dataset spanning roughly 2.8--5.2 microns at a (non-constant) resolution of $R \approx 2700$. 

Stage 3 of \texttt{Eureka!}\ performs optimal spectrum extraction \citep{Horne1986}. Stage 4 generates light curves, and as in \cite{esparza2023detection}, we opt for native \textit{pixel} binning rather than native (spectral) resolution element binning, extracting every pixel with an assigned wavelength value. The native pixel binning is expected to retain the greatest information content and the greatest sensitivity to the intrinsic CO band structure. This approach comes at the cost of a potential increase in pixel-to-pixel correlations, which we discuss in Section~\ref{sec:why_xcorr}. Stage 5 of \texttt{Eureka!} performs light curve fitting. We adopt the orbital parameters and limb darkening recommendations from \cite{carter2024benchmark}. For each channel, we model the transit with \texttt{batman} \citep{kreidberg2015batman}, fitting only for transit depth in each channel. We include a linear ramp in time and a jump step function to model the change in flux after the mirror tilt event \cite{alderson2023early}. 

As previously described, our nominal atmospheric retrievals are not performed on a typical transmission spectrum. Rather, we perform our primary atmospheric fits at the spectroscopic lightcurve level, dividing out the temporal ramp and step function in each channel (with the timing of the step function set by a fit to the white light lightcurve).  
We package the lightcurves into arrays of shape ($N_{\rm exposure}$, $N_{\rm pixel}$). We do not perform any binning along the time axis, retaining 160 exposures.
Separately, to directly compare our approach to traditional methods, we also perform secondary analyses on a more standard data product: a transmission spectrum binned to $R=100$ from the native-resolution fits discussed above. 

\subsection{The forward model}
Performing inference requires a mapping from the latent parameter space (i.e., the exoplanet atmosphere's properties) to the data space (i.e., the spectrum). To this end, we generate model transmission spectra with the \texttt{CHIMERA} code \citep{line2013systematiciii}. Our spectra include gas opacity from $\rm H_2O$ \citep{polyansky2018exomol}, $\rm CO$ \citep{li2015rovibrational}, $\rm CO_2$ \citep{yurchenko2020exomol}, and $\rm SO_2$ \citep{underwood2016exomol,tobias2018critical}. These gas species warrant inclusion because they have been firmly detected in the atmosphere of WASP-39~b \citep[e.g.,][]{rustamkulov2022analysis} and are expected based on forward modeling. We do not include methane in the list of species in our forward model because it has not been detected in this planet's atmosphere, and it is not anticipated to be a dominant carbon carrier (see discussion in Section~\ref{sec:co_met_results}). This parameterization is appropriate for a ``free retrieval'' --- we do not enforce that the volume mixing ratios of the considered gases follow predictions from equilibrium chemistry. We also incorporate opacity from Rayleigh scattering, an optically thick gray cloud deck (at a fit pressure level $P_{\rm cloud}$), and collision-induced absorption from $\rm H_2-H_2$ and $\rm H_2-He$ pairs. 

Our forward models for this cross-correlation study differ from traditional JWST retrieval forward models in a few key ways:

\begin{enumerate}
    \item We calculate our atmospheric models at very high spectral resolution ($R\simeq10^6$). We then convolve this spectrum with the approximate instrumental line-spread function and bin to data resolution \citep[using the \texttt{spectres} package to ensure flux conservation;][]{carnall2017spectres}.
    \item To ensure that we fully resolve line cores at such high spectral resolution, we extend our model domain to very low pressures ($10^{-12}$ bars).
    \item We Doppler-shift our forward model to match a velocity offset at each orbital phase.
    \item We limb-darken our spectra as a function of orbital phase. Following \cite{gandhi2022spatially}, we perform this darkening for the morning and evening limbs separately because the morning and evening limbs occult different regions on the stellar disk. That is, the atmosphere is broken into two regions, each backlit with its own limb-darkened stellar profile.
    \item We fit for the broadening of WASP-39~b's spectral lines due to solid-body rotation  by broadening our model with a convolutional kernel. This kernel accounts for the contributions of the approaching and receding limbs to the blue and red wings of the spectral lines, respectively \citep{maguire2022high}.
\end{enumerate}

We introduce another parameter, $K_p$, which is the change in planetary radial velocity ($RV$) over the course of the transit (assuming a circular orbit):

\begin{equation}
    RV = K_p\sin(2\pi \phi) + V_{\rm sys},
\end{equation}
for planetary orbital phase $\phi$ ranging from 0 (mid-transit) to 1. While we expect $V_{\rm sys}$ to coincide with the velocity of the WASP-39 system \citep[$-58.4421$~km\,s$^{-1}$;][]{mancini2018gaps}, we allow $V_{\rm sys}$ to vary as a free parameter. For each sample in the retrieval, we compare our (Doppler-shifted) atmospheric model to the flux at each exposure, as is typically done for ground-based retrievals \citep[e.g.,][]{gibson2020detection,line2021solar}.

In a departure from standard practices in cross-correlation analyses both from the ground and space \citep[e.g.,][]{brogi2019retrieving,gibson2020detection,esparza2023detection}, we preserve the shape of the spectral continuum in our cross-correlation retrievals. Given the wavelength coverage and moderate spectral resolution of this dataset, typical continuum ``removal'' procedures are likely to also remove key spectral features (such as the blue slope of the water feature in the G395 bandpass). Essentially, the continuum is difficult to define given the broadness and ubiquity of features, so fitting and removing it would subtly remove atmospheric signal. Early tests on these data supported this intuition, revealing obvious biases in retrieved temperature (pushing toward the upper edge of the prior) and chemistry (very strong departures from expected abundance ratios). We therefore conclude that preserving the spectral continuum is necessary for obtaining accurate compositional inferences.

\subsection{The likelihood function}
The comparison between model and data requires a likelihood function to assess how well the models match the data. For our likelihood function, we again depart from standard JWST retrievals, which follow the standard Gaussian likelihood function:

\begin{equation}\label{eq:standard}
    \mathcal{L} = \sum_{i=1}^{N}\bigg{(}\frac{y_i - f(x_i)}{\sigma_i}\bigg{)}^2,
\end{equation}
for spectral data $y_i$, forward model $f$ computed on each wavelength point $x_i$, and formal spectral uncertainties $\sigma_i$. We instead employ the \cite{brogi2019retrieving} likelihood:

\begin{equation}\label{eq:xcorr}
    \ln \mathcal{L} = -\frac{N}{2}\bigg{(}\ln{s_ds_m} + \ln\bigg{[}\frac{s_d}{s_m} + \frac{s_m}{s_d} - 2C\bigg{]}\bigg{)},
\end{equation}
where $N$ is the number of data points, $s_m$ and $s_d$ are the empirically computed variance of the model and data, respectively, and $C$ is the correlation coefficient. We mean-subtract both the model and the data prior to cross-correlation.\footnote{We note that mean-subtracting the spectrum before cross-correlation likely limits our ability to retrieve information regarding cloud decks.} This approach ``nulls'' the standard Gaussian likelihood function with respect to the (assumed Gaussian-distributed) uncertainty, $\sigma$, substituting $\sigma$ for its maximum-likelihood estimator. The mapping essentially fits for the noise by performing error inflation across the spectral axis for each phase. It is otherwise mathematically equivalent to the standard Gaussian likelihood function in Eq.~\ref{eq:standard} \citep[see, e.g.,][]{gibson2020detection}.

\subsection{The sampling}
The likelihood function can be thought of as a ``cost function'' to optimize, given the constraints of the prior distribution. Calculating the likelihood function in full (e.g., on a grid) is intractable given the runtime of a single model evaluation and the high dimensionality of the retrieval problem. Retrievals therefore couple forward models to samplers that aim to efficiently explore parameter space.

While \texttt{CHIMERA} is generally coupled to \texttt{PyMultiNest} \citep{buchner2014x} to perform the statistical sampling for relevant retrievals, we choose instead to couple \texttt{CHIMERA} to the \texttt{nautilus} sampling package \citep{lange2023nautilus}. We do so because the statistical literature indicates the \texttt{MultiNest} algorithm \citep{feroz2009multinest} provides underestimated posterior widths on the order of 10\% and biased Bayesian evidence on the order of ln(Z) of 1 unless extremely conservative settings are used \citep{albert2020jaxns,nelson2020quantifying,miller2021radius,ih2021understanding,lemos2023robust,dittmann2024notes,group2025comparison}. \texttt{nautilus} has been used in the context of atmospheric retrievals \citep{gebhard2023inferring}; we benchmark its performance, confirm \texttt{MultiNest}'s biases, and demonstrate \texttt{nautilus}' substantial increase in efficiency over \texttt{MultiNest} (per live point) in Appendix~\ref{sec:sampling_benchmarking}.

\begin{table}
\centering
\begin{tabular}{ll}
\hline
Parameter & Prior \\
\hline
\multicolumn{2}{l}{\textit{Gaseous absorbers}} \\
$\rm \log_{10}{^{12}C^{16}O}$ [dex] & $\mathcal{U}(-12, 0)$ \\
$\rm \log_{10}H_2O$ [dex] & $\mathcal{U}(-12, 0)$ \\
$\rm \log_{10}CO_2$ [dex] & $\mathcal{U}(-12, 0)$ \\
$\rm \log_{10}SO_2$ [dex] & $\mathcal{U}(-12, 0)$ \\
$\rm log_{10}(^{13}C^{16}O / ^{12}C^{16}O)$ [dex] & $\rm \mathcal{U}(-5, 0)$ \\
\hline
{\textit{Thermal structure}} \\
$T$ [K] & $\mathcal{U}(200, 5000)$ \\
% $\alpha_\mathrm{scale}$ & $\mathcal{U}(0.5, 2)$ \\
\hline
% \multicolumn{2}{l}{\textit{Stellar Contamination Parameters}} \\
% $F_\mathrm{spot}$, $F_\mathrm{fac}$ & Triangle point picking, $\sum F_i = 0.5$ \\
% $T_\mathrm{phot}$ [K] & $\mathcal{U}(3000, 7000)$ \\
% $T_\mathrm{spot}$ [K] & $\mathcal{U}(3000, T_\mathrm{phot})$ \\
% $T_\mathrm{fac}$ [K] & $\mathcal{U}(T_\mathrm{phot}, 7000)$ \\
% \hline
{\textit{Velocity}} \\
$V_{\rm rot}$ [km/s] & $\mathcal{U}(0.1, 10)$ \\
$V_{\rm sys}$ [km/s] & $\mathcal{U}(-73.4421, -43.4421)$ \\
$K_{p}$ [km/s] & $\mathcal{U}(-300, 300)$ \\
% NRS1-NRS2 Offset [ppm] & $\mathcal{N}(0, 40)$ \\
\hline
{\textit{Miscellaneous}} \\
$\mathrm{x}R_p$ & $\mathcal{U}(0.5, 2.0)$ \\
\hline
\end{tabular}
\caption{Prior distributions for our retrieval parameters. $\mathcal{U}(a,b)$ denotes a uniform distribution between $a$ and $b$. %, while $\mathcal{N}(\mu,\sigma)$ denotes a normal distribution with mean $\mu$ and standard deviation $\sigma$. 
 Samples in which the gas abundances sum to greater than unity are rejected.} %Temperature priors follow a hierarchical structure where $T_\mathrm{spot} \leq T_\mathrm{phot} \leq T_\mathrm{fac}$.}
\label{table:priors}
\end{table}

\begin{figure}
    \centering
    \includegraphics[scale=0.3]{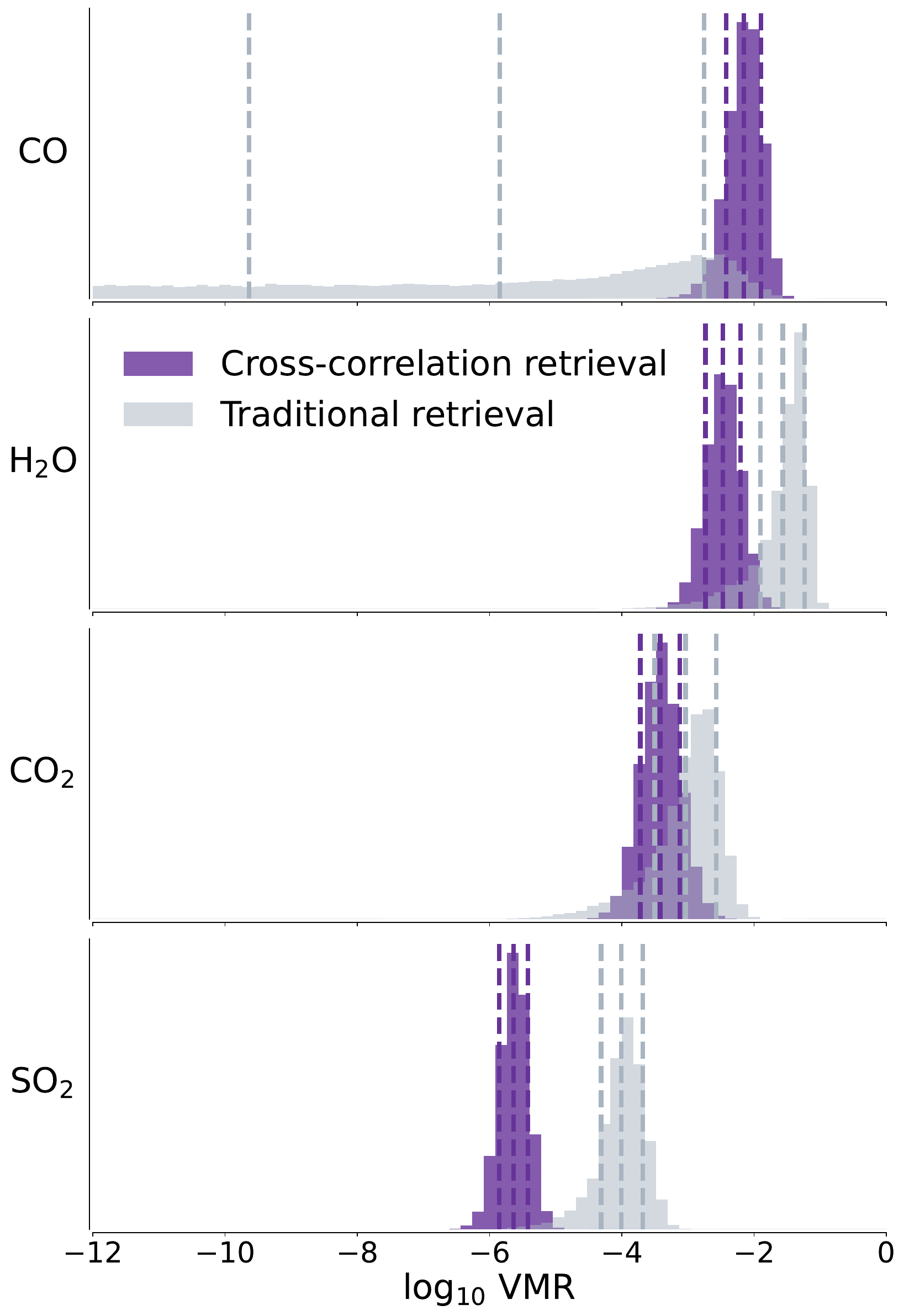}
    \caption{Our native pixel resolution cross-correlation constraints on gases in WASP-39~b using G395H data vs.\ those acquired with standard methods ($R=100$, phase-integrated). Our method is sensitive to previously hidden gases' abundance. Specifically, we obtain a bounded abundance for CO.  Vertical dashed lines mark the 16th and 84th percentiles for each distribution.}
    \label{fig:methods_compare}
\end{figure}

We present our prior distribution ranges in Table~\ref{table:priors}. For the gaseous absorbers, abundances are each drawn from a log-uniform distribution between $-12$ and 0 dex. We fit for the abundance of the $\rm ^{13}C^{16}O$ isotopologue \textit{relative} to the abundance of $\rm ^{12}C^{16}O$, as we expect our retrieval to be sensitive to the \textit{ratio} of these species' sharply wavelength-dependent opacity structures \citep[e.g.,][]{gandhi2023retrieval}. We place the additional constraint that the minor $\rm ^{13}C^{16}O$ abundance  must lie within 5 dex of $\rm ^{12}C^{16}O$, generously encompassing the carbon isotope fractionation from the Solar System \citep[$\rm \log_{10}{^{12}C^{16}O} - \log_{10}{^{13}C^{16}O} \approx 1.9$;][]{clayton2004astrophysics}, the interstellar medium \citep[$\rm \log_{10}{^{12}C^{16}O} - \log_{10}{^{13}C^{16}O} \approx 1.8$;][]{milam200512c}, and planetary-mass companions \citep[$\rm \log_{10}{^{12}C^{16}O} - \log_{10}{^{13}C^{16}O} \approx 1.5$;][]{line2021solar}. We allow the isothermal atmospheric temperature to widely bracket the equilibrium temperature of WASP-39~b \citep[1170~K;][]{mancini2018gaps}. We also fit a scaling factor to the planetary radius, x$R_p$, fixing the reference pressure to 1~bar.

As for the velocity parameters, our priors allow a very wide range of planetary rotational velocities (0.1 to 10~km\,s$^{-1}$) that bracket the expected velocity under synchronous rotation ($\approx 1.6$~km\,s$^{-1}$). We allow systemic velocities ($V_{\rm sys}$) within 15~km\,s$^{-1}$ of the reported systemic velocity \citep[$\approx -58~\mathrm{km\,s^{-1}}$;][]{mancini2018gaps}. This restricted prior width avoids self-aliasing of the CO cross-correlation function \citep[e.g.,][]{esparza2023detection} while remaining more than three orders of magnitude wider than the uncertainty distribution of the system's measured velocity \citep{mancini2018gaps}. We freely vary $K_p$ over a wide range (\rm $-300$--$300~\mathrm{km\,s^{-1}}$) covering the expected value under a circular orbit \citep[$\approx 129~\mathrm{km\,s^{-1}}$)][]{mancini2018gaps}; we do not expect this value to be well constrained given how slowly this planet moves and the relative low resolution of the observing mode.

We perform evidence testing to assess the significance of our detections of individual species. That is, we perform additional retrievals without individual molecules, then compare the Bayesian evidence of each retrieval to the full ``reference'' model to assess the significance of each gas's inclusion in our model \citep[e.g.,][]{benneke2013,welbanks2026challenges}.

\section{Results} \label{sec:results}
We now discuss the results of our phase-resolved, cross-correlation retrievals on native pixel resolution data. We address the gas abundance and velocity parameters first, then present the inferred distributions of C/O and atmospheric metallicity. Our full retrieval corner plots can be found in Appendix~\ref{sec:full_corners}.

\subsection{Gas abundances and velocity parameters}\label{sec:velocity_abunds}

Our retrieved gas abundances are presented in Fig.~\ref{fig:methods_compare}. As with previous analyses of this dataset \citep{alderson2023early}, we find that the abundance of both the major and minor CO isotopologues are essentially unconstrained using traditional analysis techniques (gray contours).  Our $R=100$ transmission spectrum retrieval does not retrieve a bounded CO abundance, and the gas is preferred with only $\Delta \ln(Z) = 0.2$. 

In contrast, we obtain a bounded CO abundance ($-2.1^{+0.2}_{-0.3}$; purple contours) with the cross-correlation likelihood function applied to the native pixel resolution, phase-resolved spectrum. This retrieved abundance is consistent with the much less informative constraint provided by traditional retrieval techniques and from other works in the literature that combine spectra from multiple JWST modes (see Section~\ref{sec:compare}).  Notably, our retrieved CO abundance lies near the upper bound associated with the non-detection using traditional methods. 
The presence of CO is strongly preferred in our cross-correlation analysis, with $\Delta \ln(Z) = 23$.

Our cross-correlation retrieval also finds bounded constraints on the gas-phase abundances of $\rm H_2O$ ($-2.5\pm0.3$), $\rm CO_2$ ($-3.4 \pm0.3$), and $\rm SO_2$ ($-5.6\pm0.2$). Tests with a non-isothermal thermal profile \citep[][not shown]{guillot2010radiative} also yielded consistent results. These constraints agree well with expectations from the literature \citep{constantinou2023early,kawahara2025differentiable}. They are also consistent with our  $R=100$ retrievals within $\approx2\sigma$ for each gas except $\rm SO_2$. For the latter, the discrepancy is intriguing, but it does not impact our results on metallicity and C/O due to its low intrinsic abundance. This is the first cross correlation detection of SO$_2$ in the literature, and its interpretation merits future work beyond the scope of this Letter.  Overall, for all gases other than CO, our cross-correlation retrievals achieve comparable precision to the $R=100$ retrievals. 

For water in particular, we retrieve a slightly lower abundance than with traditional methods, despite using the same opacities in both cases to fit the same underlying data set.  In this case, we believe the cross-correlation retrieval to be more accurate because we probe many individual water spectral lines with this technique; whereas the standard approach only fits a single \textit{partial} water band and is therefore susceptible to detector edge systematics as well as degeneracies with aerosol parameters.  

Beyond H$_2$O, given that all of our retrieved abundances, with the exception of CO, are lower with the cross correlation retrieval, we also hypothesize that at lower resolution CO is degenerate with the abundances of the rest of the gaseous species.  If this is the case, the other gases must increase in abundance to fit the CO opacity function in the low-resolution retrieval, whereas this degeneracy is resolved at native pixel resolution. However, additional forward modeling and mock retrieval work beyond the scope of this Letter would be required to confirm this hypothesis.

Our velocity parameters are not tightly constrained but are consistent with their expected values.  Notably, our $V_{\rm sys}$ measurement is strongly bimodal (see Fig.~\ref{fig:corner_xcorr_isotherm}).  We find that this bimodaility is strongly correlated with the CO isotopologue abundance. The bluer mode is likely the ``correct'' mode for two reasons. First, it is consistent with the known systemic velocity of \mbox{WASP-39} within 1$\sigma$. Second, this bluer mode correlates with a plausible value for the minor CO isotopologue's relative abundance (consistent with the interstellar medium). The redder velocity mode prefers a wide range of very low minor isotopologue abundances, implying that this bimodality is due to a spectral correlation between the major and minor CO isotopologue. Interestingly, this indicates that the G395H data do have some sensitivity to the minor CO isotopologue, in line with the findings of \citet{esparza2023detection}. Our $K_p$ constraint is unbound, with a spread of $\approx 100$~km\,s$^{-1}$.

The velocity constraint on $V_{\rm sys}$ also appears to be driven by the presence of CO. When we remove the opacity of both CO isotopologues from our retrieval, the constraint on $V_{\rm sys}$ becomes unbounded. Furthermore, when we fit a separate $V_{\rm sys}$ for the NRS1 and NRS2 detectors, the $V_{\rm sys}$ for NRS1 pushes to the red edge of the prior, while the $V_{\rm sys}$ for NRS2 (which contains the bulk of the CO opacity in the G395 bandpass) converges to a similar result as our retrieval with a single $V_{\rm sys}$ for both detectors. This result implies that the velocity constraint in our retrieval is driven by the comb-like structure of the CO opacity.

\subsection{C/O and metallicity}\label{sec:co_met_results}
\begin{figure}
    \centering
    \includegraphics[width=1\linewidth]{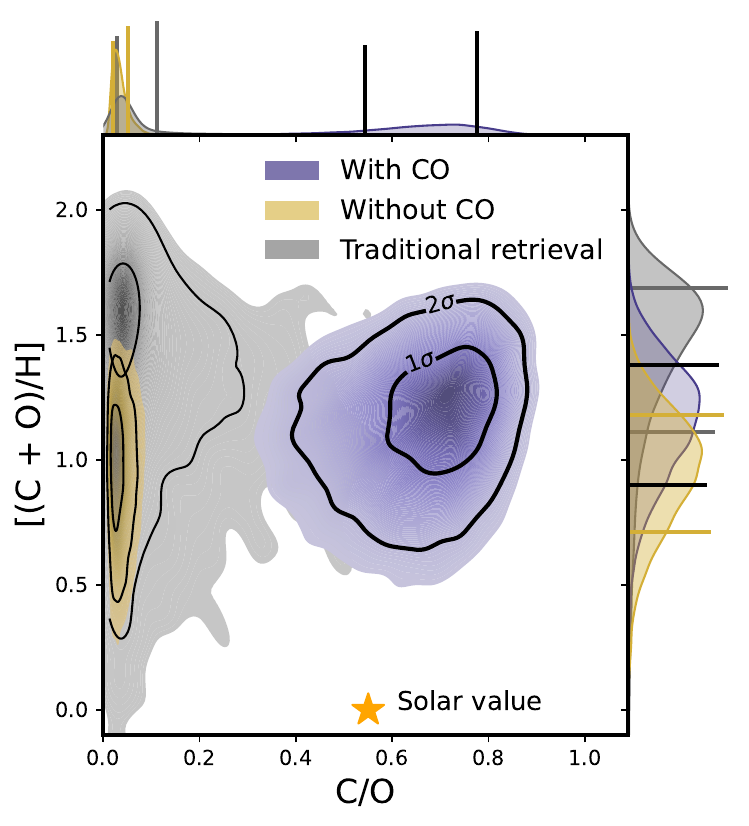}
    \caption{The (volatile-proxy) envelope metallicity and C/O ratio of WASP-39~b, as inferred by our native pixel resolution, cross-correlation, phase-resolved free retrieval (purple). With a single NIRSpec/G395H transit, we are able to simultaneously constrain both chemical parameters.  As points of comparison, we also show the posterior distributions obtained without including the contribution of CO abundance (yellow) and using standard JWST retrieval techniques (gray).  1$\sigma$ and 2$\sigma$ contours are outlined in black for all three posterior distributions.  The orange star indicates the solar values. The ticks on the marginals mark the 16th and 84th quantiles of each distribution.}
    \label{fig:metallicity_c_to_o}
\end{figure}

We use our freely retrieved molecular abundances to construct estimates of the volatile-proxy metallicity (i.e., [(C+O)/H]) and C/O ratio of the gas-phase envelope (Fig.~\ref{fig:metallicity_c_to_o}). The metallicity is constrained to be \mbox{$15^{+9}_{-6}$ $\times$ solar}, and the C/O is constrained to be $0.68^{+0.10}_{-0.14}$. This result is broadly consistent with other estimates from the literature based on JWST data for WASP-39~b, as we discuss in greater detail in Section~\ref{sec:compare}.

Compared to the cross-correlation retrieval, our retrieval using traditional techniques obtains a consistent (but somewhat higher) metallicity (\mbox{$30^{+18}_{-18}$ $\times$ solar}).
However, because it does not constrain CO, the latter finds a much lower C/O ($0.04^{+0.08}_{-0.01}$). These distributions are shown in gray contours in Fig.~\ref{fig:metallicity_c_to_o}.  We can see from Fig.~\ref{fig:methods_compare} that the low inferred C/O is presumably an artifact of the non-detection of the CO molecule at low spectral resolution, combined with somewhat higher retrieved H$_2$O and SO$_2$ abundances.  In particular, the CO non-detection pulls down the median of the CO posterior distribution by 2 orders of magnitude relative to the bounded abundance measured by the cross-correlation retrieval.  Because CO is the dominant carbon reservoir in WASP-39~b, its non-detection has an outsized impact on the inferred C/O.  This effect can also be seen in the yellow contours in Fig.~\ref{fig:metallicity_c_to_o}, where we report the metallicity and C/O from our cross-correlation retrieval, but neglecting the contribution from CO gas, resulting in a comparably low C/O.

We note that $\rm CH_4$ is not included in our retrieval setup because previous studies have shown no evidence for this molecule in WASP-39~b \citep[e.g.,][]{rustamkulov2022analysis,alderson2023early}, despite JWST's strong leverage to detect CH$_4$ at high confidence. Furthermore, to alter our C/O or metallicity constraints by more than 1\% would require
$\rm CH_4$ to be present at a mixing ratio of  \mbox{$>-3$ dex}. Such high abundances of methane are strongly disfavored in other retrievals that have been performed on JWST data \citep[e.g.,][]{kawahara2025differentiable}.

\section{Discussion} \label{sec:discussion}
\subsection{Comparison to other work}\label{sec:compare}

Firstly, our results agree with \cite{esparza2023detection}, \cite{grant2023detection}, and \cite{esparza2025testing}: at higher spectral resolution, NIRSpec/G395H provides substantial evidence for the presence of CO. Our results build on this statement by providing a quantitative \textit{measurement} of the CO abundance using these same archival data. Our retrievals do not reveal a bounded constraint on the minor carbon isotope due to a bimodal correlation between this parameter and $V_{\rm sys}$. In the velocity mode corresponding to the systemic velocity of WASP-39, the minor CO isotopologue has an abundance $-1.6^{+0.8}_{-1.3}$ relative to the major isotopologue, indicating sensitivity to this species. Even so, Bayesian odds ratios do not indicate that we have formally detected it ($\Delta\ln Z < 1$).

Our analysis is also broadly consistent with other studies that have applied retrievals to binned JWST data of WASP-39~b, both in the G395H bandpass \citep{niraula2023origin,wilkinson2024breaking, kawahara2025differentiable} and in other instrument modes, such as NIRSpec PRISM \citep{constantinou2023early,khorshid2024constraining,roy2025role}, NIRISS/SOSS \citep{fisher2024jwst,changeat2025cloud}, PRISM and SOSS combined \citep{constantinou2024vira}, MIRI LRS \citep{powell2024sulfur}, NIRCam, SOSS, PRISM, and G395H separately \citep{lueber2024information}, and NIRISS, NIRCam, NIRSpec, and MIRI together \citep{ma2025new}. Specifically, these studies and our own agree in inferring an elevated metallicity and low-to-moderate C/O (both relative to solar) for WASP-39~b.  Our retrieved C/O using cross-correlation techniques trends toward the highest values reported in the literature, and we justify that this inference has improved accuracy based on the bounded abundance measurement of CO.

To more directly compare our retrieval results to an array of traditional retrievals, we turn to the outputs of the JWST Transiting Exoplanet Community Early Release Science Program (JTEC-ERS) Model Synthesis effort \citep[Welbanks et al., in prep;][]{welbanks_synthesis}. This study aims to establish best practices for atmospheric retrievals in the JWST era by benchmarking a number of retrieval codes run by different research groups against a single dataset: the panchromatic JWST spectrum of WASP-39~b spanning 0.5--5.5 $\mu$m. Our results are consistent within 2$\sigma$ with each molecules' free retrieval average from the Model Synthesis effort, and our results are consistent with the Model Synthesis metallicity within 0.3$\sigma$.\footnote{L. Welbanks, private communication.}

\subsection{What drives the signal quality?}
\subsubsection{Why does native pixel resolution perform well?}\label{sec:why_xcorr}
Our results confirm that higher spectral resolution provides more leverage on the CO band structure. In the limit of very low spectral resolution, the CO opacity ``comb'' becomes smeared out; in the limit of very high spectral resolution, these features are very clearly differentiable from a flat line and the opacity structure of other competing gaseous species. Therefore, while the standard retrieval yields substantial probability mass at moderately high CO mixing ratios, it finds that the CO pseudocontinuum is not necessary to explain the data (hence the formal non-detection).

The cost of extracting native pixel resolution spectra is the enhanced pixel-to-pixel correlations in these data. The cross-correlation likelihood presented in Eq.~\ref{eq:xcorr} was developed specifically for this regime, in which the observed flux may not be well described with well-known Gaussian uncertainties. Additional tests that we performed seem to align with this intuition: traditional retrievals performed at native pixel resolution using the Gaussian likelihood function did not yield bounded CO constraints. The question of the ideal resolution for traditional retrievals is an intriguing one that merits exploration in future work.

\subsubsection{Why does phase-resolving perform well?}
As discussed, WASP-39~b subtends a non-negligible range of orbital phase during its transit. Its transmission spectrum is therefore Doppler-shifted as a function of time. This effect should be especially pronounced for species such as CO, given how narrow its spectral features are at native pixel resolution. Accounting for this Doppler-shifting by comparing atmospheric models directly to lightcurves --- and appropriately shifting the atmospheric models themselves --- is a straightforward solution. This approach is routinely employed in high-resolution ground-based exoplanet studies \citep[e.g.,][]{brogi2019retrieving}, and more recently to a handful of JWST phase curves and secondary eclipses \citep{sing2024absolute, ouyang2026cross,snellen2026cancri}. We have shown here that it extends well to JWST NIRSpec/G395H transits, even in the limit of slower-moving exoplanets.

\section{Conclusion} \label{sec:conclusion}
This Letter provides the proof of concept that an exoplanet's atmospheric C/O ratio and metallicity can be measured robustly with the JWST NIRSpec/G395H instrument mode, alone. This result is made possible by applying techniques that had previously been exclusively employed by the ground-based observing community --- i.e., phase-resolved cross-correlation retrievals on native pixel resolution spectra. In the case of WASP-39~b's NIRSpec/G395H data, our method yields additional sensitivity to the presence of CO compared to standard techniques.  We confidently detect the gas and produce a bounded abundance measurement.  We simultaneously measure bounded abundances for all of the key carbon- and oxygen-bearing species present in \mbox{WASP-39 b's} atmosphere. 

Our results firmly demonstrate that traditional approaches that bin JWST data in wavelength and in time can dilute information about exoplanet atmospheric composition. We also show that the non-detection of CO, which is a common occurrence in NIRSpec/G395 datasets, can heavily bias inferred C/O ratios (as well as metallicities to a lesser extent) by ``hiding'' the dominant carbon-carrying molecule in hot Jupiter atmospheres.
Applying the high-resolution phase-resolved retrieval techniques from this Letter to additional G395H observations of transiting exoplanets is likely to be fruitful, providing a pathway to maximally make use of legacy datasets.

\restartappendixnumbering

\begin{contribution}
ABS designed and performed the analysis and drafted the manuscript. EMRK advised on the experimental design and manuscript preparation. EM reduced the data for this work. All authors provided feedback on initial project framing and the final manuscript.
\end{contribution}

\begin{acknowledgments}
We thank Hinna Shivkumar for interesting conversations regarding inter-pixel correlations and stellar contamination. We also thank Måns Holmberg for useful discussions on noise sources and likelihood functions. Finally, we thank Megan Bedell and the Flatiron CCA Astro Data Group for useful discussions about spectrograph performance.
ABS and EMRK acknowledge funding from the NASA Exoplanet Research Program through grant No.\ 80NSSC25K7154. JPW acknowledges support from the Swiss National Science Foundation (SNSF) under grant 10002706 (PIs: D. Kitzmann, H.J. Hoeijmakers) and from the Canadian Space Agency (CSA) under grant 24JWGO3A-03.
This work benefited from lectures, tutorials, and discussions at the ``Hi-Res in the Desert'' workshop, held at Arizona State University in December 2025, supported by NSF AAG grants 2307177 and 2307178.
The authors acknowledge the University of Maryland supercomputing resources (\url{http://hpcc.umd.edu}) made available for conducting the research reported in this paper.
This research has made use of the Astrophysics Data System, funded by NASA under Cooperative Agreement 80NSSC21M0056.
This research has made use of the NASA Exoplanet Archive, which is operated by the California Institute of Technology, under contract with the National Aeronautics and Space Administration under the Exoplanet Exploration Program.

\end{acknowledgments}

\software{\texttt{astropy} 
\citep{astropy:2013, astropy:2018, astropy:2022}, \texttt{CHIMERA} \citep{line2013systematic},  \texttt{FastChem} \citep{Fastchem2018, stock2022fastchem}, \texttt{GitHub Copilot} \citep{chen2021evaluating}, \texttt{Jupyter} \citep{granger2021jupyter}, \texttt{Matplotlib} \citep{Hunter:2007}, \texttt{nautilus} \cite{lange2023nautilus}, \texttt{Numpy} \citep{harris2020array}, \texttt{pandas} \citep{mckinney-proc-scipy-2010, reback2020pandas}, \texttt{PLATON} \citep{zhang2019forward,zhang2020platon,zhang2025retrievals}, \texttt{PyMultiNest} \citep{buchner2014x}, \texttt{scipy} \citep{2020SciPy-NMeth}, \texttt{spectres} \citep{carnall2017spectres}, 
\texttt{tqdm} \citep{da2019tqdm}}

\appendix
\renewcommand{\thefigure}{A\arabic{figure}}

\setcounter{figure}{0}

\section{Benchmarking nested sampling}\label{sec:sampling_benchmarking}

As mentioned previously, in this work we couple \texttt{CHIMERA} to \texttt{nautilus} because it provides benefits over the more standard \texttt{PyMultiNest} sampler. The primary functional difference between \texttt{nautilus} and \texttt{PyMultiNest} is the way in which the two codes define iso-likelihood surfaces as they sweep through parameter space. \texttt{PyMultiNest} defines ellipsoidal bounds and inflates the bounds by some $\epsilon$; \texttt{nautilus} learns the iso-likelihood surface by training an ensemble of neural networks.

Due to the relative novelty of applying \texttt{nautilus} in the context of atmospheric retrievals, we seek to benchmark its performance in simulated test cases. We test the performance of \texttt{nautilus} and MultiNest along three axes: posterior accuracy, evidence accuracy, and runtime / efficiency.\footnote{All code for these tests was run on a 2021 MacBook Pro with a 16~GB Apple M1 Pro chip.} To keep this problem tractable, we do not perform these tests on the full, nominal JWST retrievals presented in this manuscript. Rather, we test the performance of \texttt{nautilus} by coupling it to the \texttt{PLATON} forward model \citep{zhang2019forward,zhang2020platon,zhang2025retrievals}. Our benchmark case is the transmission spectrum of HD~209458b as observed with Hubble/STIS, Hubble/WFC3, and Spitzer \citep{knutson2007using, deming2013infrared, evans2015uniform}. We fit this with an 8-parameter model: stellar radius, planet mass, planet radius, planetary (isothermal) temperature, a log scattering slope factor, a log planet metallicity, a log (gray) cloud top pressure, and a scalar multiplier to the observed flux uncertainties.

\begin{figure}
    \centering
    \includegraphics[width=0.78\linewidth]{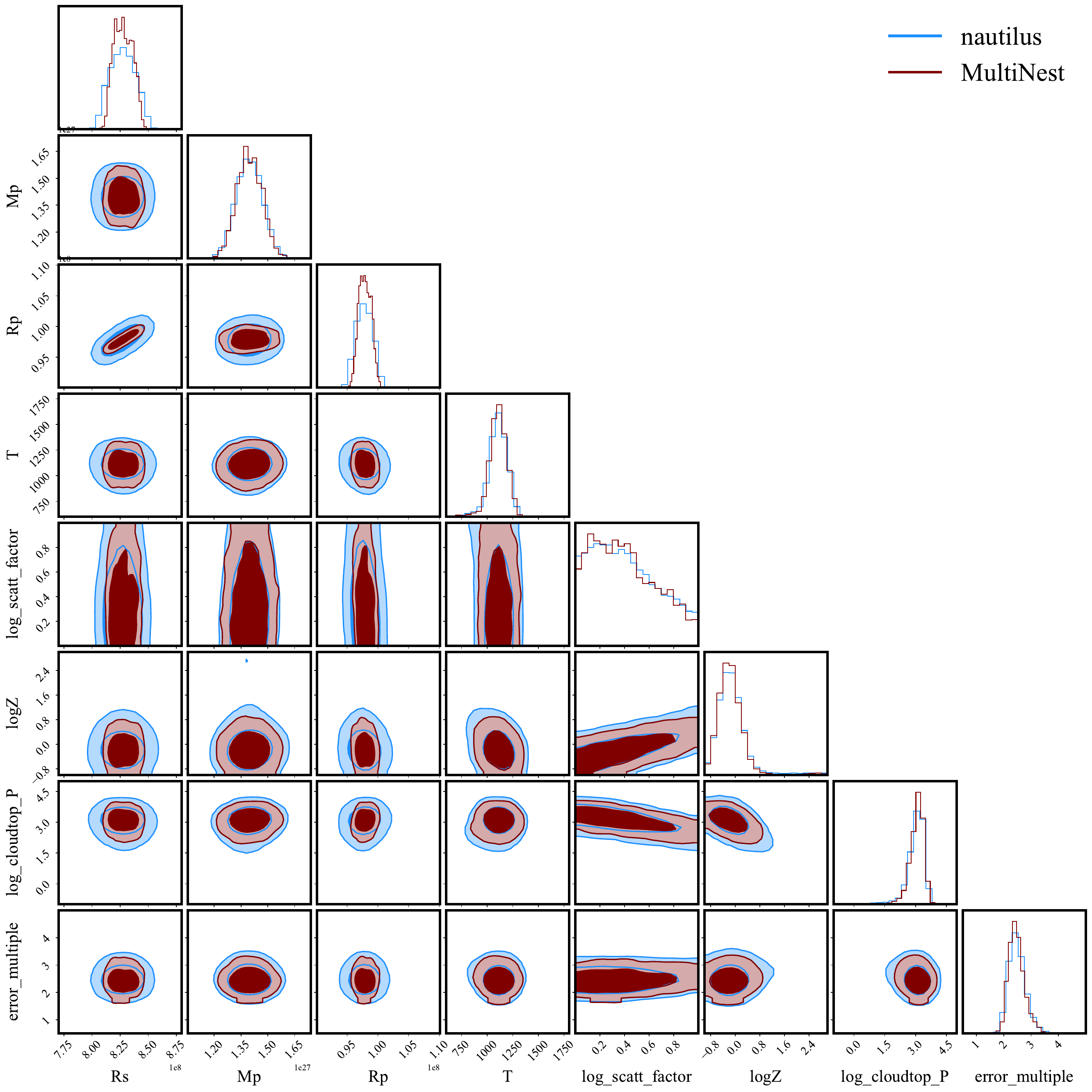}
    \caption{The posterior distributions output by \texttt{nautilus} and \texttt{MultiNest} for our \texttt{PLATON} benchmarking exercise.}
    \label{fig:nautilus_comparison_posterior}
\end{figure}

Our baseline calculations include 500 live points for \texttt{MultiNest} and a sampling efficiency of 0.3, per the code's documented suggestions for calculating Bayesian evidence. For \texttt{nautilus}, we use the developer's recommended setting of 2000 live points. We additionally set the \texttt{n\_networks} parameter to 16. We do so because the cost of training more neural networks to learn the iso-likelihood ellipsoids is negligible compared to the cost of our likelihood calculation (about 1~s per forward model computation).

\subsection{Posterior precision}
Fig.~\ref{fig:nautilus_comparison_posterior} shows the posterior distributions calculated by both \texttt{nautilus} and \texttt{MultiNest}. We find that the \texttt{MultiNest}-derived posterior distribution widths tend to be underestimated on the order of 10\% compared to the \texttt{nautilus} runs. This underestimation is particularly evident by eye in the 2D credible regions, in which differences in the joint posterior distributions (including correlations) become apparent. There does not appear to be a correlation between the simplicity of a marginalized distribution (e.g., how similar it is to a 1-dimensional Gaussian) and how well \texttt{MultiNest} matches the \texttt{nautilus} output. This underestimation is consistent with the findings of, e.g., \cite{ih2021understanding} and \cite{lemos2023robust}.

\subsection{Evidence accuracy}

\begin{figure}
    \centering
    \includegraphics[width=0.34\linewidth]{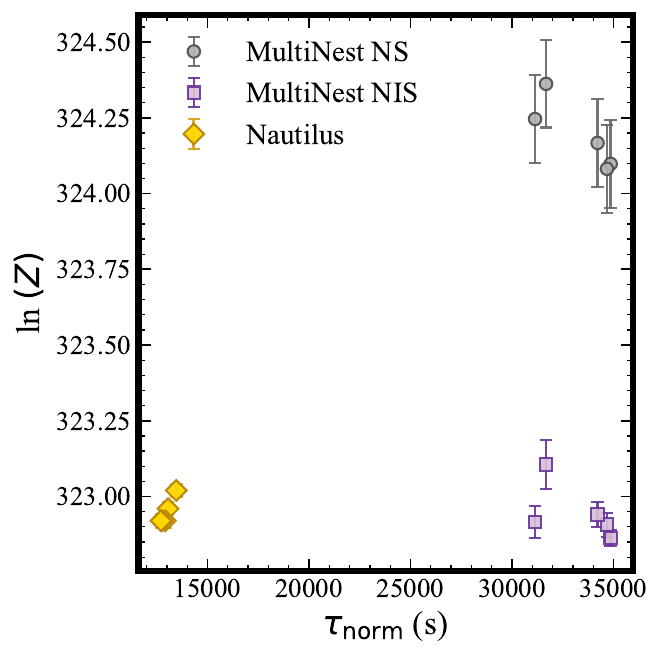}
    \caption{The result of benchmarking the \texttt{nautilus} code against \texttt{pymultinest}. The estimated natural log of the Bayesian evidence is shown as a function of normalized runtime ($\tau_{\rm norm}=2000\tau_{\rm runtime} /  n_{\rm live points}$). The nested sampling (NS) evidence and nested importance sampling (NIS) evidence as calculated by \texttt{MultiNest} are both shown.}
    \label{fig:nautilus_comparison_evidence}
\end{figure}

We find that the standard evidence as reported by \texttt{MultiNest} is inaccurate and imprecise (Fig.~\ref{fig:nautilus_comparison_evidence}), as also noted by \citet{dittmann2024notes},  \citet{lemos2023robust}, and \citet{albert2020jaxns}, likely due to the way in which \texttt{MultiNest} draws ellipsoidal bounds. While \texttt{MultiNest}'s importance nested sampling evidence is indeed more accurate and more precise than the standard reported evidence \citep{feroz2013importance}, we find that its uncertainty is underestimated by a factor of approximately 1.5 with respect to the \texttt{nautilus} output. This effect is either a systematic offset or a hint of greater variance between runs.

Notably, our benchmarking test may be in the regime in which \texttt{MultiNest} comparatively performs slightly \textit{better} than the novel retrievals presented in this Letter. \cite{albert2020jaxns} found that \texttt{Multinest}'s evidence bias rapidly grows with the number of dimensions past 10 dimensions. The test problem in this appendix is 8-dimensional; the full retrievals we present in this manuscript reach 11 dimensions. We expect that any evidence biases would therefore be elevated in the context of our scientific problem. Given that the bias is expected to be dimension- and problem-dependent \citep{albert2020jaxns}, this evidence bias will likely result in Bayes factor biases, as well.

\subsection{Runtime / efficiency}
Despite having four times as many live points as our \texttt{MultiNest} runs, \texttt{nautilus} only took roughly two times as long to converge on average (Fig.~\ref{fig:nautilus_comparison_evidence}). Scaled by the number of live points, then, \texttt{nautilus} is far more efficient --- as found in general test cases by \cite{lange2023nautilus}.

\section{Full retrieval outputs}\label{sec:full_corners}

\begin{figure}
    \centering
    \includegraphics[width=0.95\linewidth]{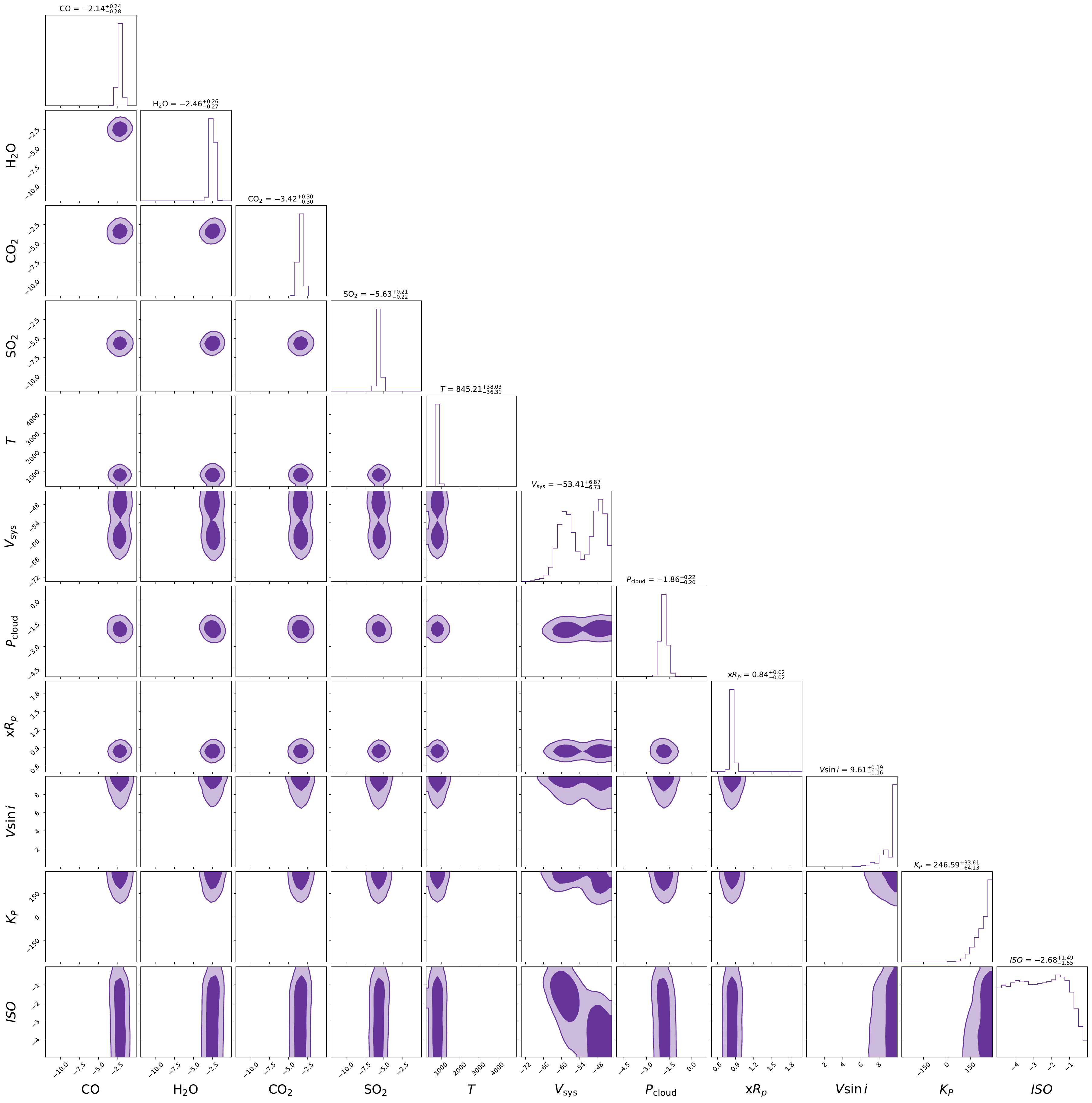}
    \caption{Posterior distribution of our cross-correlation retrievals, with gases and cloud deck pressure fit in $\log_{10}$. We label the relative log abundance of the minor CO isotopologue as ``ISO.''}
    \label{fig:corner_xcorr_isotherm}
\end{figure}

\begin{figure}
    \centering
    \includegraphics[width=0.84\linewidth]{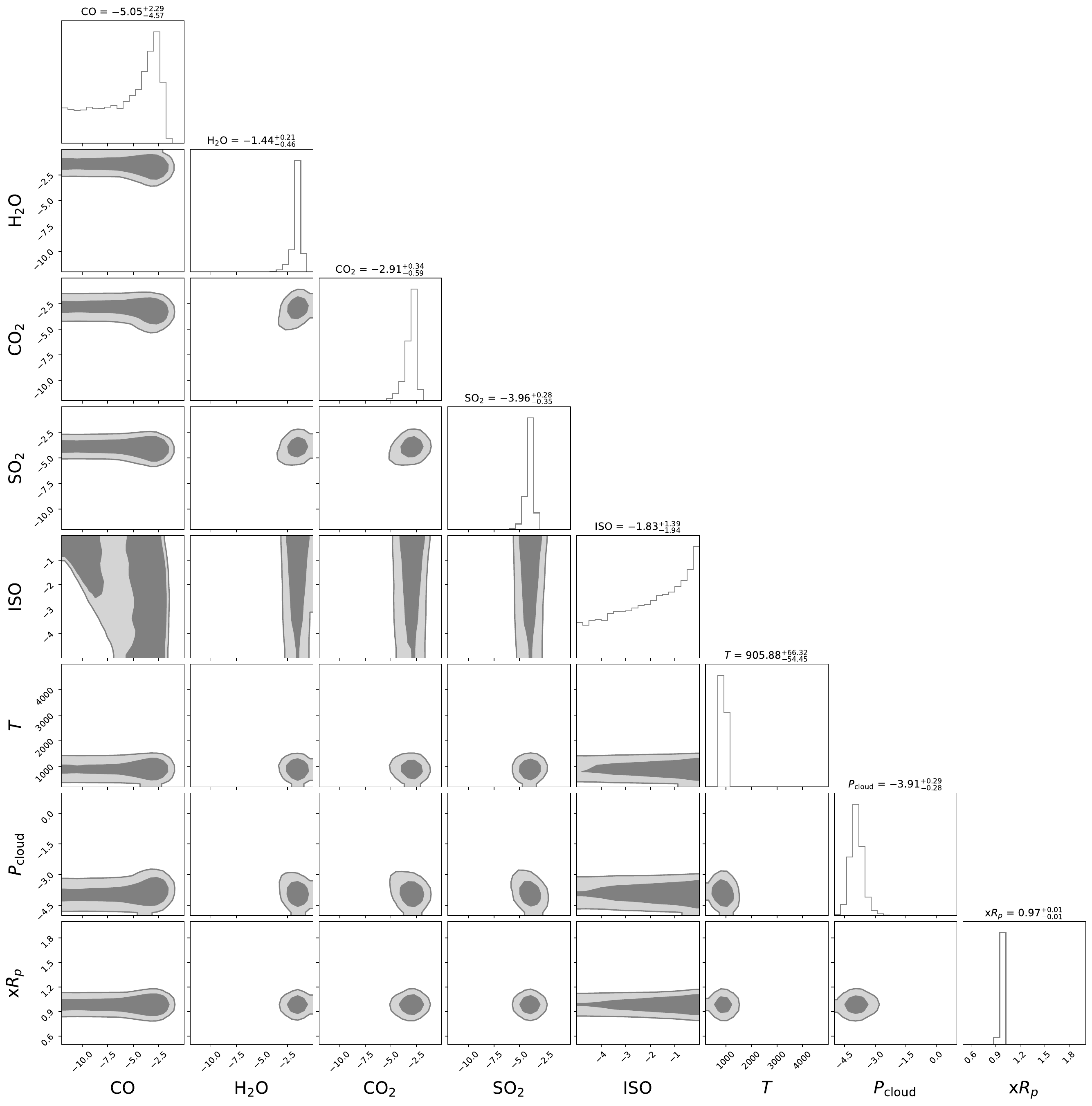}
    \caption{Posterior distribution of our traditional retrievals.}
    \label{fig:corner_standard_isotherm}
\end{figure}

Our full retrieval results will be made available on Zenodo upon publication.  Full corner plots for our cross correlation retrieval and traditional retrieval are shown in Figs.~\ref{fig:corner_xcorr_isotherm} and \ref{fig:corner_standard_isotherm}, respectively.

\bibliography{sample631}{}
\bibliographystyle{aasjournalv7}



\end{document}